\pdfoutput=1
\documentclass{turing2012}

\usepackage{graphicx}
\usepackage{amsmath, amsthm, amssymb}
\usepackage{overpic}

\theoremstyle{definition}
\newtheorem{definition}{Definition}

\newcommand{\minus}{\scalebox{0.5}[1.0]{$-$}}

\begin{document}

\title{Exploiting Vagueness for Multi-Agent Consensus}
\author{
Michael Crosscombe \and Jonathan Lawry\\
Department of Engineering Mathematics, \\University of Bristol, \\BS8 1UB, United Kingdom\\\
\texttt{m.crosscombe@bristol.ac.uk $\cdot$ j.lawry@bristol.ac.uk}
}

\maketitle

\begin{abstract}
A framework for consensus modelling is introduced using Kleene's three valued logic as a means to express vagueness in agents' beliefs.
Explicitly borderline cases are inherent to propositions involving vague concepts where sentences of a propositional language may be \emph{absolutely true}, \emph{absolutely false} or \emph{borderline}.
By exploiting these intermediate truth values, we can allow agents to adopt a more vague interpretation of underlying concepts in order to weaken their beliefs and reduce the levels of inconsistency, so as to achieve consensus. We consider a consensus combination operation which results in agents adopting the borderline truth value as a shared viewpoint if they are in direct conflict. Simulation experiments are presented which show that applying this operator to agents chosen at random (subject to a consistency threshold) from a population, with initially diverse opinions, results in convergence to a smaller set of more precise shared beliefs. Furthermore, if the choice of agents for combination is dependent on the payoff of their beliefs, this acting as a proxy for performance or usefulness, then the system converges to beliefs which, on average, have higher payoff.

\hspace{1em}


\end{abstract}

\section{Introduction}
\label{sec:1}
Reaching a consensus by agreeing a shared viewpoint or position is a fundamental part of many multi-agent decision making and negotiation scenarios. In this paper we argue that by exploiting vagueness in the form of explicitly \emph{borderline} cases we can define an operator for belief combination which not only allows a population of agents to reach consensus but also results in them adopting, on average, a more \emph{useful} set of beliefs. The basic intuition underlying this operator is that conflicting agents can agree to allocate borderline truth values to propositions about which they hold inconsistent beliefs. For example, two individuals, one of which believes that `Cameron is an effective prime minister' whilst the other believes that `Cameron is ineffective', may agree, in some circumstances, to adopt the shared view that `Cameron is borderline effective/ineffective'.

Of course, beliefs about the world do not exist in isolation but inform and influence our decisions and actions. From this perspective, some sets of beliefs are more positive or useful than others, resulting in better long term performance, perhaps by making the individuals concerned richer, happier or just better able to survive. More generally, in a multi-agent context, different beliefs result in different actions, collecting different payoffs. In this paper we present simulation studies which show that implementing our proposed operator across a population of agents, initially holding diverse beliefs, results in convergence to a smaller subset of more precise shared opinions. Furthermore, under the assumption that better performing agents, i.e. those with higher payoff, are more likely to interact to reach consensus, we show that the range of beliefs obtained at steady state are on average better, i.e have higher payoff, than the agents' initial beliefs. The formalism adopted here is that of Kleene's three valued logic and the operator investigated has been proposed for single propositions in \cite{perron} and extended to multi-propositional languages in \cite{lawrydubois}.

An outline of the paper is as follows: Section \ref{sec:2} gives a brief overview of consensus modelling. Section \ref{sec:3} introduces Kleene logic and the three valued consensus combination operator. Section \ref{sec:4} describes simulation experiments in which agents are selected at random to form a consensus provided that they are sufficiently consistent with one another. In section \ref{sec:5} we introduce a payoff function for beliefs, so that the payoff of a particular set of beliefs acts as a proxy for the performance of an agent holding those beliefs. We then adapt the experiments described in section \ref{sec:4} so that the probability of an agent being selected for consensus is proportional to their payoff. Finally, in section \ref{sec:6} we give some discussions and conclusions.

\section{Background and Related Work}
\label{sec:2}
A number of models for consensus have been proposed in the literature which have influenced the development of the framework described in this paper.
\cite{degroot} introduced a model for reaching a consensus involving a weighted, global updating of beliefs, iterating until an agreement is reached.
In DeGroot's model, agents assign a weight distribution to the population before forming a new opinion.
By applying their assigned weights to the other agents' beliefs, an agent can control the influence that others have on their own beliefs.

As an alternative to DeGroot's model, the Bounded Confidence (BC) model introduced in \cite{krause98} provides agents with a confidence measure.
An agent quantifies their level of confidence in their own opinions and are then able to limit their interactions to those agents who possess similar beliefs if they are highly confident (small bounds), or extend the range of possible interactions if the agents possess low confidence (large bounds). In this model agents do not a priori assign weights to the beliefs of others, but instead determine such weightings based on similarity and on their own confidence levels. This is similar in essence to the inconsistency threshold that we introduce in section \ref{sec:3}, but applied on an individual basis.

The Relative Agreement (RA) model \cite{deffuant02} then extends the Bounded Confidence model to allow agents to assign weights to the beliefs of others by quantifying the extent of the overlap of their respective confidence bounds. By having agents declare a confidence interval for their beliefs, the model then restricts interactions to those pairs of agents with overlapping intervals. Consequently, agents are only required to assess their own beliefs and are not required to make explicit judgements about those of other agents. \cite{deffuant02} also moved to a model of pair-wise interactions to better capture social interactions of individuals, the latter being a setting in which group-wide updates to beliefs are unintuitive in that they do not reflect typical social behaviour.

A fundamental difference between our approach and the above models is that we use Kleene's three valued logic to represent beliefs in a propositional logic setting, rather than identify opinions with real values or intervals.
\cite{perron} have shown that through use of a three-state model for networked consensus of complete graphs, nodes converge to a consensus much faster and with greater accuracy when compared to a restrictive binary model. In the sequel we extend this approach to a more general setting involving larger languages and incorporating a measure of payoff for beliefs.

\section{A Three Valued Consensus Model}
\label{sec:3}
In this section we introduce Kleene's three valued logic \cite{kleene52} as a model of explicitly borderline cases resulting from the inherent vagueness of propositions. We adopt a propositional logic setting as follows: Let ${\cal L}$ be a finite language of propositional logic with connectives $\wedge$, $\vee$ and $\neg$, and propositional variables ${\cal P}=\{p_1, \ldots, p_n \}$. Also, let $S{\cal L}$ denote the sentences of ${\cal L}$ generated by recursively applying the connectives to the propositional variables in the usual manner. A Kleene valuation then allocates truth values $0$ (false), $\frac{1}{2}$ (borderline) and $1$ (true) to the sentences of ${\cal L}$ as follows:

\begin{definition}{Kleene Valuations}

		A Kleene valuation $v$ on  $\mathcal{L}$ is a function $v : S\mathcal{L} \rightarrow \{0, \frac{1}{2}, 1\}$ such that $\forall \theta, \varphi \in S\mathcal{L}$ the following hold:

		\begin{itemize}
			\item $v(\neg \theta) = 1 - v(\theta)$

			\item $v(\theta \land \varphi) = $ \textnormal{min}$(v(\theta),v(\varphi))$

			\item $v(\theta \lor \varphi) = $ \textnormal{max}$(v(\theta), v(\varphi))$
		\end{itemize}
The truth table for Kleene valuations are shown in table \ref{ttk}.
	\label{KV}
\end{definition}

\begin{table}
\centering
\begin{tabular}{|c||c|c|} \hline
$\neg$ & $1$ & $0$ \\ \hline
 & $\frac{1}{2}$ & $\frac{1}{2}$ \\ \hline
  & $0$ & $1$ \\ \hline
\end{tabular} \hspace*{0.25cm}
\begin{tabular}{|c||c|c|c|} \hline
$\wedge$ & $1$ & $\frac{1}{2}$ & $0$ \\ \hline \hline
$1$ & $1$ & $\frac{1}{2}$ & $0$ \\ \hline
$\frac{1}{2}$ & $\frac{1}{2}$ & $\frac{1}{2}$ & $0$ \\ \hline
$0$ & $0$ & $0$ & $0$ \\ \hline
\end{tabular}
\hspace*{0.25cm}
\begin{tabular}{|c||c|c|c|} \hline
$\vee$ & $1$ & $\frac{1}{2}$ & $0$ \\ \hline \hline
$1$ & $1$ & $1$ & $1$ \\ \hline
$\frac{1}{2}$ & $1$ & $\frac{1}{2}$ & $\frac{1}{2}$ \\ \hline
$0$ & $1$ & $\frac{1}{2}$ & $0$ \\ \hline
\end{tabular}
\caption{Kleene truth tables.}
\label{ttk}
\end{table}

It is sometimes convenient to represent a Kleene valuation $v$ by its associated \emph{orthopair} $(P,N)$ \cite{lawrydubois}, where $P=\{p_i \in {\cal P}: v(p_i)=1\}$ and $N=\{p_i \in {\cal P}: v(p_i)=0 \}$. Notice that $P \cap N = \emptyset$ and that $(P \cup N)^c$ corresponds to the set of borderline propositional variables.

Kleene valuations have been proposed as a suitable formalism in which to capture explicitly borderline cases as resulting from inherent flexibility in the definition of vague concepts in natural language \cite{lawrytang,lawrygonzalez}. For example, consider the proposition `Ethel is short'. For the concept short, we might identify a lower height threshold $\underline{h}$ below which any height is classed as being \emph{absolutely short}, and similarly there may be an upper threshold $\overline{h}$ above which any height is \emph{absolutely not short}. If Ethel's height lay between $\underline{h}$ and $\overline{h}$ then this would result in a borderline truth value for the statement `Ethel is short'.

It is important to note that the middle truth value $\frac{1}{2}$ is not intended to represent epistemic uncertainty, but rather explicitly borderline cases resulting from the inherent vagueness of natural language propositions. Hence, if we say that the statement `Ethel is short' is borderline true/false we are not saying that the truth or falsity of this proposition is unknown. Instead we are indicating that Ethel's height is a borderline case of the predicate short. In order to emphasise the difference between the epistemic and the borderline interpretation of three valued logic it is helpful to think in terms of conditioning. For instance, if we learn that it is \emph{unknown} whether or not Ethel is short, then this provides us with no new information about her height. In contrast, learning that Ethel is \emph{borderline short} does provide us with new information about Ethel's height, namely that it lies on the borderline between short and not short. A more comprehensive discussion of these issues can be found in \cite{ciucci}. A consequence of using this interpretation of the middle truth value is that in the current paper we only model consensus for sets of propositions which admit borderline cases.  In other words, our approach can be used for propositions such as `Ethel is short' but not, for example, for the proposition `Ethel is strictly less that 1.4 metres tall'.


The following three valued consensus operator was described in detail in \cite{lawrydubois}:

\begin{definition}{Consensus Operator}
\label{consensus_operator}

Let $v_1$ and $v_2$ be Kleene valuations on ${\cal L}$ with associated orthopairs $(P_1,N_1)$ and $(P_2,N_2)$. Then the consensus $v_1 \odot v_2$ is the Kleene valuation with the orthopair
	\begin{align*}
		 ((P_1 \minus N_2) \cup (P_2 \minus N_1), (N_1 \minus P_2) \cup (N_2 \minus P_1))
 	\end{align*}
\end{definition}
The corresponding truth table for this operator is shown in table \ref{tab:consensus_op}. From this we can see that the operator preserves the non-borderline truth values $0$ or $1$ except in the case of a direct conflict i.e. when one agent has truth value $1$ and the other $0$. In this case both agents adopt the middle truth value $\frac{1}{2}$. Alternatively, from definition \ref{consensus_operator} we can think of $\odot$ as an operator which initially weakens both opinions so as to remove direct inconsistencies, before then combining them.

\begin{table}
\parbox{.45\linewidth}{
	\centering
	\begin{tabular}{| r || c | c | c | c |} 
		\hline 
		\small{$\odot$}& \small{$1$} & \small{$\frac{1}{2}$} & \small{$0$}  \\   [0.5ex]
		\hline \hline
		\small{$1$} & \small{$1$} & \small{$1$} & \small{$\frac{1}{2}$} \\ 
		\hline
		\small{$\frac{1}{2}$} & \small{$1$} & \small{$\frac{1}{2}$} & \small{$0$} \\
		\hline
		\small{$0$} & \small{$\frac{1}{2}$} & \small{$0$} & \small{$0$} \\ [1ex] 
		\hline
	\end{tabular}
	\caption{Truth table for the consensus operator.}
	\label{tab:consensus_op}
}
\hfill
\parbox{.45\linewidth}{
	\centering
	\begin{tabular}{| r || c | c | c | c |} 
		\hline 
		\small{$I$}& \small{$1$} & \small{$\frac{1}{2}$} & \small{$0$}  \\   [0.5ex]
		\hline \hline
		\small{$1$} & \small{$0$} & \small{$0$} & \small{$1$} \\ 
		\hline
		\small{$\frac{1}{2}$} & \small{$0$} & \small{$0$} & \small{$0$} \\
		\hline
		\small{$0$} & \small{$1$} & \small{$0$} & \small{$0$} \\ [1ex] 
		\hline
	\end{tabular}
	\caption{Inconsistency truth table.}
	\label{tab:inconsistency}
}
\end{table}

We now introduce two measures that will be used throughout the subsequent simulation experiments.

\begin{definition}{A Measure of Vagueness}
\label{vagueness}

Let $v$ be a Kleene valuation on ${\cal L}$ with orthopair $(P,N)$ and $n$ propositional variables. Then we measure the vagueness of $v$ by the proportion of propositional variables which it classifies as being borderline. That is:
	\begin{align*}
		V(v) = \frac{\left|(P\cup N)^c\right|}{n}
	\end{align*}

\end{definition}

\begin{definition}{Inconsistency Measure}
\label{inconsistency}

Let $v_1$ and $v_2$ be Kleene valuations on ${\cal L}$ with corresponding orthopairs $(P_1,N_1)$ and $(P_2, N_2)$. Then we define the inconsistency measure of $v_1$ and $v_2$ to be the proportion of propositional variables which are in direct conflict between the two valuations i.e. $v_1(p_i)\neq \frac{1}{2}$,  $v_2(p_i)\neq \frac{1}{2}$ and $v_1(p_i)=1-v_2(p_i)$. That is:
\begin{align*}
		I(v_1, v_2) = \frac{\left|(P_1\cap N_2)\right| + \left|(P_2\cap N_1)\right|}{n}
	\end{align*}
\end{definition}

Table \ref{tab:inconsistency} shows the inconsistency truth table of two valuations for a propositional variable, highlighting the cases where two valuations are \textsl{inconsistent}, and consistent otherwise.
We can see that there is a probability of $\frac{2}{9}$ that two valuations will be inconsistent for each propositional variable in the language.
In the sequel we will propose a threshold $\gamma \in [0,1]$ on inconsistency so that valuations $v_1$ and $v_2$ can be combined only if $I(v_1,v_2)\leq \gamma$.

\section{Simulation Experiments based on Random Selection of Agents}
\label{sec:4}

\begin{figure*}[t]
	\centering
	\includegraphics[width=0.8\textwidth]{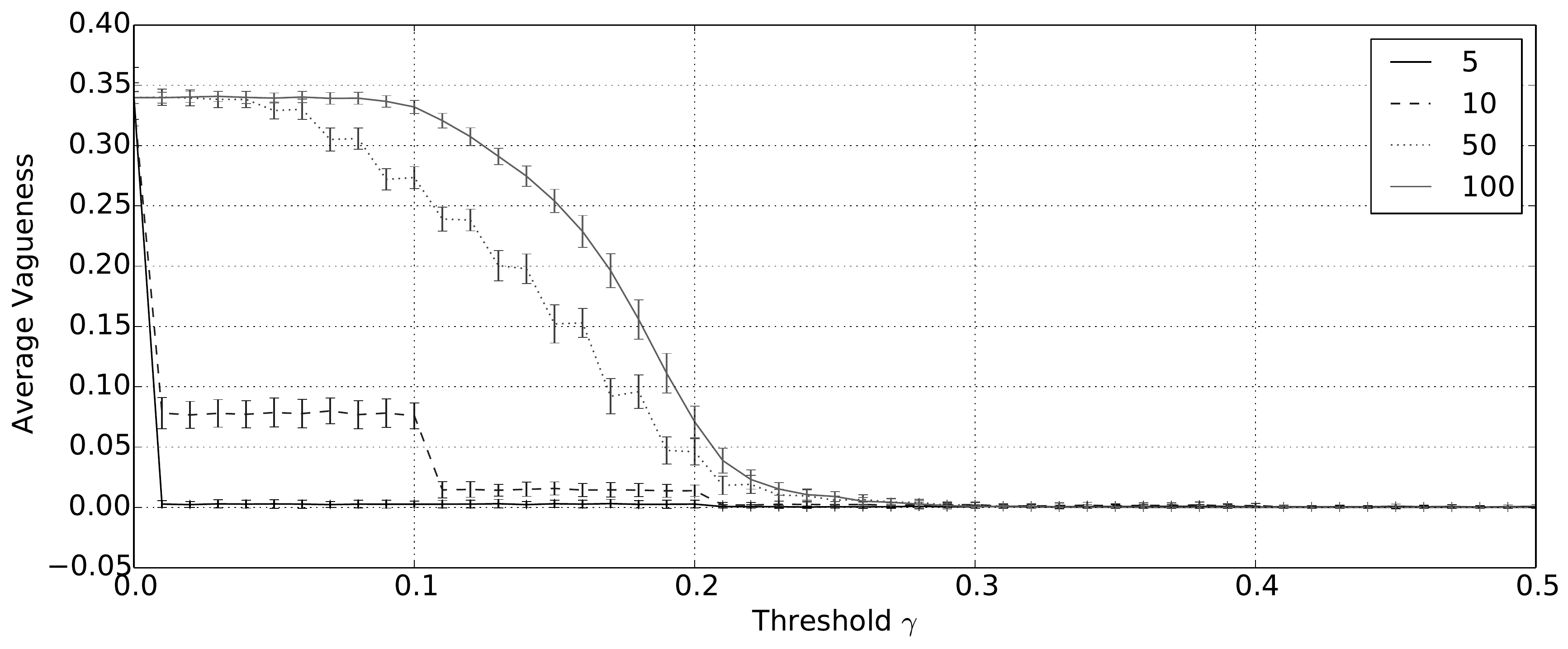}
	\caption{ Average vagueness after $50,000$ for varying inconsistency thresholds $\gamma$ and language sizes.}
	\label{fig:vagueness_vagueness}
	\vspace{1em}
	\includegraphics[width=0.8\textwidth]{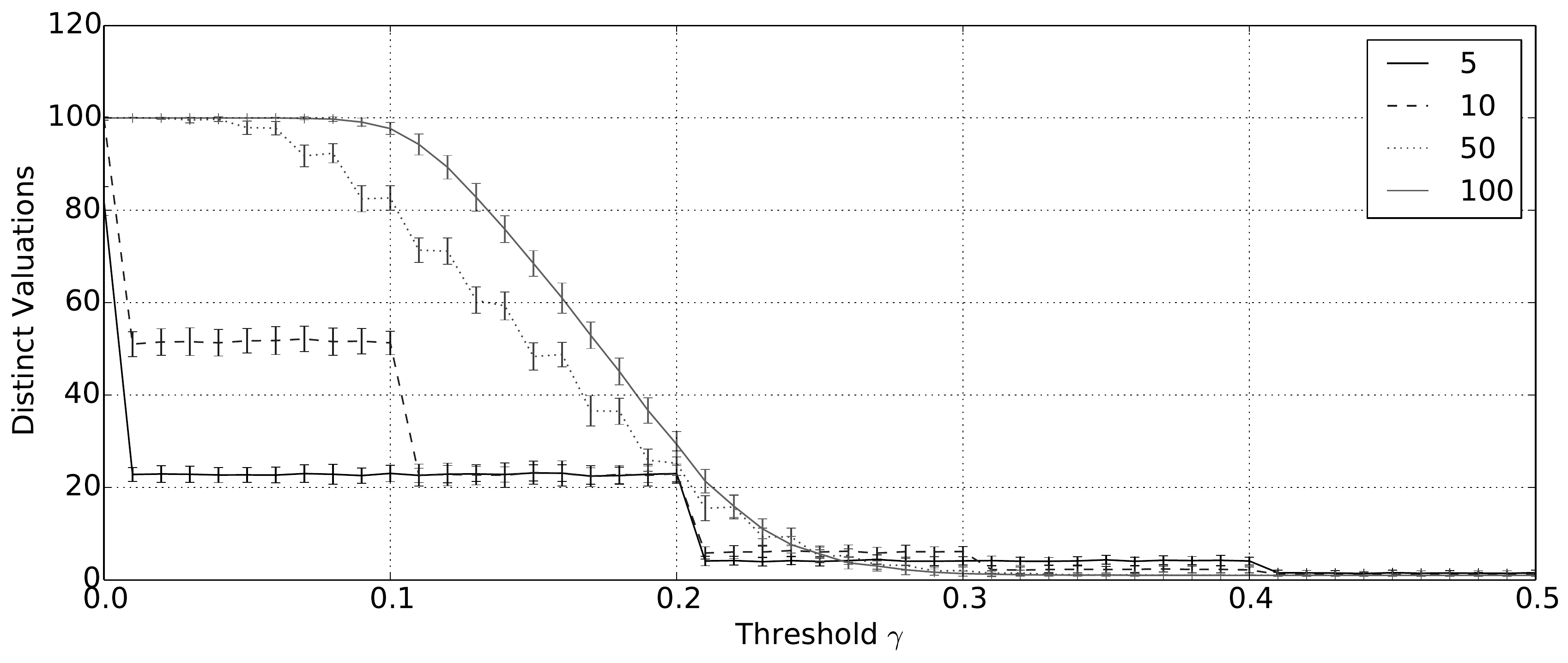}
	\caption{ Number of distinct valuations after $50,000$ iterations for varying inconsistency thresholds $\gamma$ and language sizes.}
	\label{fig:vagueness_dvps}
\end{figure*}

We introduce simulation experiments in order to investigate the convergence properties of the three valued logic operator when implemented across a multi-agent system. The experimental set up is loosely based on those proposed in \cite{deffuant02} and \cite{meadows12}, although our representation of opinions is quite different with beliefs taking the form of Kleene valuations on ${\cal L}$, rather than vectors of bounded real numbers.

We will consider two distinct initialisations of the beliefs of a population of agents. The \emph{random three valued initialisation} allocates the truth values $0$, $\frac{1}{2}$ and $1$ to each agent and each propositional variable at random i.e. with probability $\frac{1}{3}$ for each truth value. In contrast, the \emph{random Boolean initialisation} only allocates the binary truth values $0$ and $1$, each with a probability of $\frac{1}{2}$. This latter initialisation will be required in section \ref{sec:5} in order to directly compare the proposed three valued combination operator with a similar two valued operator. In this section we will use the random three valued initialisation in order to investigate the extent to which the three valued operator results in convergence to a shared set of opinions across the population of agents.

We set a fixed maximum number of $50,000$ iterations\footnote{In preliminary experiments we found that $50,000$ was an upper bound on the number of iterations required for the system to reach steady state across a range of parameter settings.}. At each time step a pair of agents are selected at random from the population. An inconsistency threshold value $\gamma \in [0,1]$ is set, so that for any pair of agents with respective valuations $v_1$ and $v_2$, if $I(v_1,v_2) \leq \gamma$ then both agents replace their beliefs with the consensus valuation $v_1 \odot v_2$, while if $I(v_1,v_2)>\gamma$ then no combination is performed and both agents retain their original beliefs. For $\gamma=1$ we obtain what is equivalent to the totally connected graph model described in \cite{perron}, in which any pair of agents can combine their beliefs, whilst taking $\gamma=0$ corresponds to the most conservative scenario in which only absolutely consistent beliefs can be combined. The parameters for the simulation experiments are then as follows:
\begin{itemize}
		\item Population size: $100$
		\item Language size i.e $|{\cal P}|=n$: $5, 10, 50, 100$
		\item Initial beliefs: Random three valued.
		\item Inconsistency threshold: $\gamma \in [0,1]$.
\end{itemize}

Figures \ref{fig:vagueness_vagueness} and \ref{fig:vagueness_dvps} show the results for the experiments after $50,000$ iterations. In each case the plots show mean values with error bars representing standard deviation across $100$ independent runs of the simulation. Figure \ref{fig:vagueness_vagueness} shows the average vagueness determined by taking the mean value of $V(v)$ (definition \ref{vagueness}) across the population. Note that for a random three valued initialisation of beliefs we expect a mean vagueness value of $\frac{1}{3}$ at the start of the simulation. As the threshold $\gamma$ increases then the average vagueness decreases to zero, so that for $\gamma \geq 0.3$ we are left with almost entirely crisp (i.e. Boolean) opinions. In general the more conservative the combination rules (i.e. requiring higher levels of consistency) then the more it is that vague beliefs are maintained in the population. Figure \ref{fig:vagueness_dvps} shows the number of distinct valuations (i.e. different opinions) remaining in the population after $50,000$ iterations. Again this decreases with $\gamma$ and for $\gamma > 0.4$ agents have on average converged to a single shared belief. This is consistent with the analytical results presented in \cite{perron} for the single propositional, $\gamma=1$ case.

\section{Simulation Experiments Incorporating a Payoff Model}
\label{sec:5}

In this section we extend the simulation framework described in section \ref{sec:4} to allow for different payoffs for different beliefs. As outlined in section \ref{sec:1}, payoff is introduced as a proxy for performance, and is motivated by the intuition that different beliefs result in different actions which then, over time, lead to different levels of performance. Here we adopt an abstract simplification of this process in which each Kleene valuation is allocated a real valued payoff. Then, instead of being selected at random for combination, an agent is picked from the population according to a probability which is proportionate to the payoff value of their beliefs.
The idea, then, is that agents with better or more useful opinions will be more successful and furthermore, it will be these successful agents who will be most likely to need to reach a consensus between them.

Here the underlying intuition is that, in real systems it is the most successful agents, with the highest payoff values, who are most likely to find themselves in conflict with one another, and who will most benefit from reaching an agreement. We adopt a simple summative payoff model in which each propositional variable $p_i$ is allocated a value in the range $[-1,1]$, denoted $f(p_i)$, and the payoff for a valuation with orthopair $(P,N)$ is then calculated  as  follows:
	\begin{align*}
		f(P,N) = \sum_{p_i \in P} f(p_i) - \sum_{p_i \in N} f(p_i)
	\end{align*}
Another perspective on this type of payoff function is as follows: For each propositional variable $p_i$, a truth value of $1$ results in a payoff $f(p_i)$ (which can be either positive or negative), a truth value of $0$ results in the opposite signed payoff $\minus f(p_i)$, and a borderline truth value $\frac{1}{2}$ results in a neutral payoff of $0$. The payoff value for a Kleene valuation $v$ is then simply taken to be the sum of the payoffs for each propositional variable under the truth values allocated by $v$.
\begin{center}
	\begin{figure*}[t]
		\centering
		\includegraphics[width=0.8\textwidth]{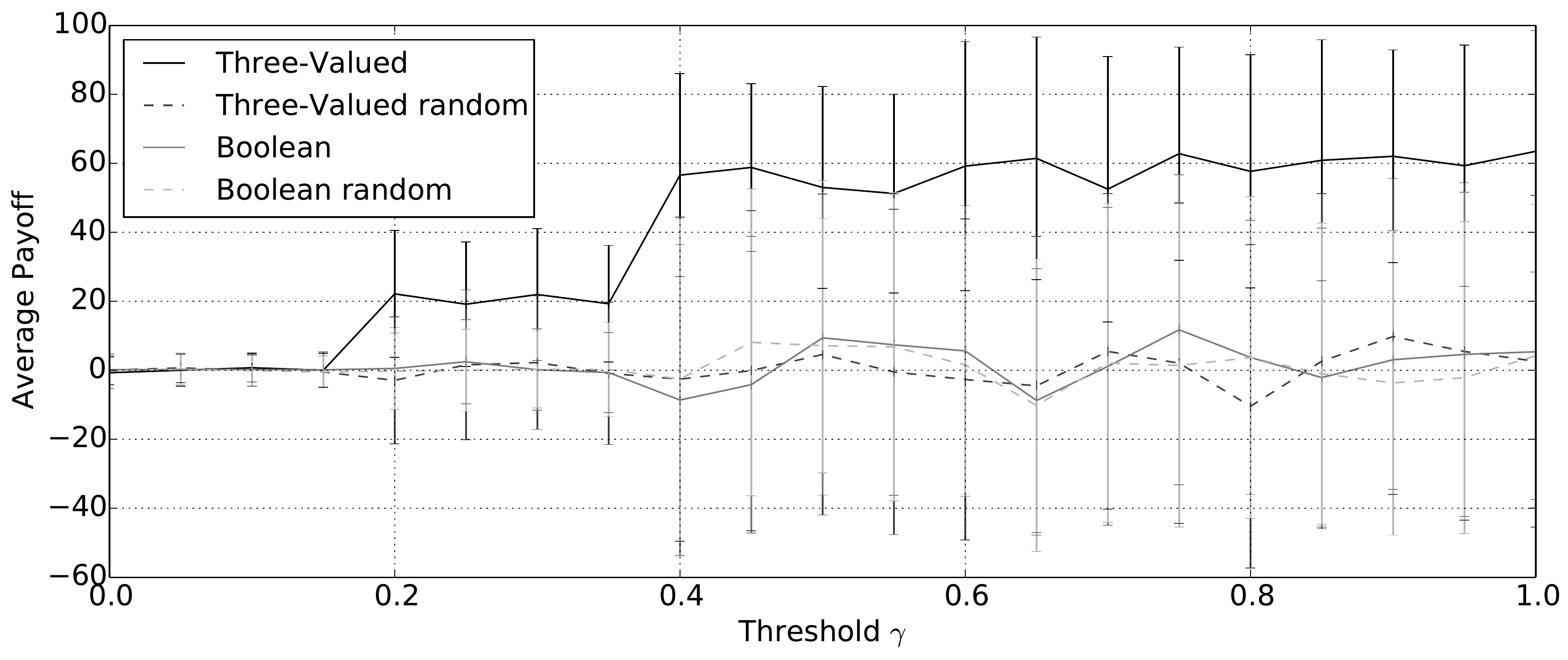}
		\caption{ Average payoff after $50,000$ iterations for varying inconsistency thresholds $\gamma$, shown as a percentage of the maximal payoff.}
		\label{fig:average_payoff}
		\vspace{1em}
		\includegraphics[width=0.8\textwidth]{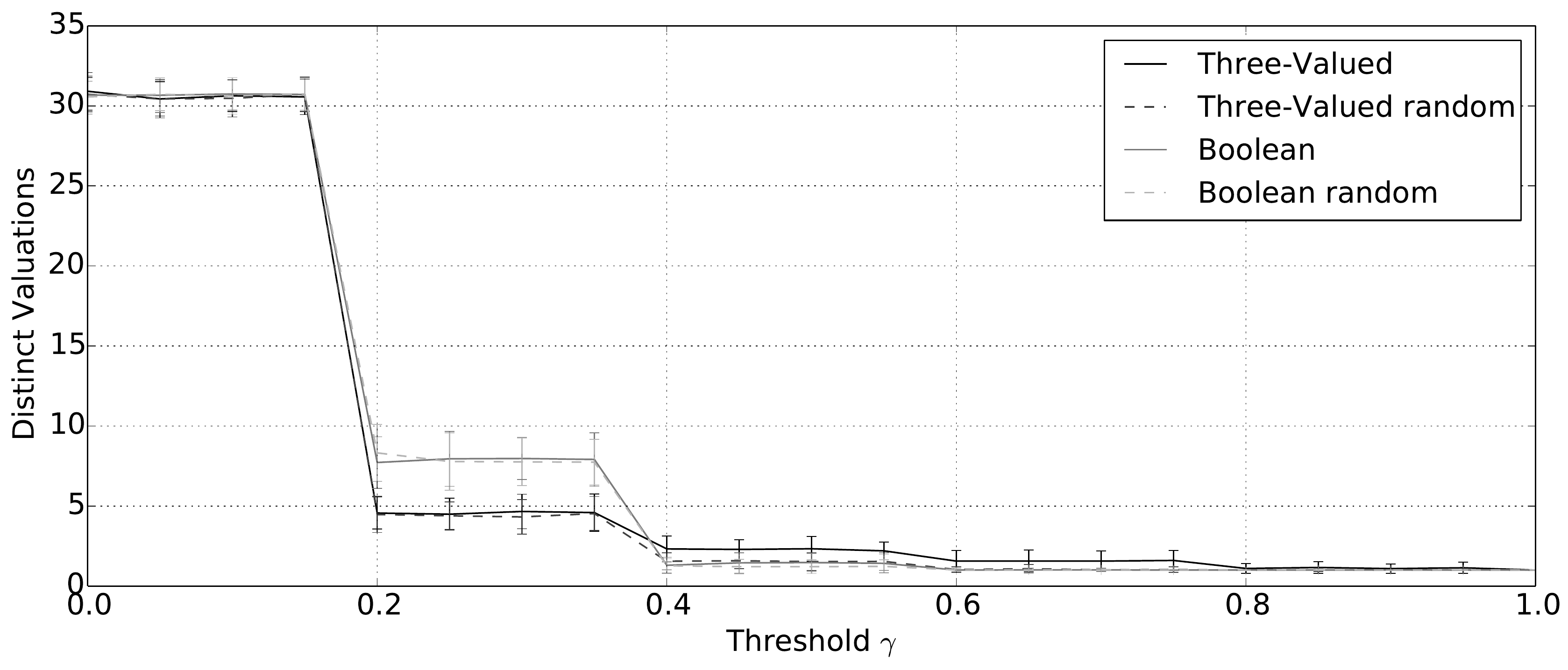}
		\caption{ Number of distinct  valuations  after $50,000$ iterations for varying inconsistency thresholds $\gamma$.}
		\label{fig:payoff_dkvs}
	\end{figure*}
\end{center}

Based on payoff values we define a probability distribution over the agents in the population according to which the probability that an agent with beliefs $(P,N)$ is selected for possible consensus combination is proportional to $f(P,N)+n$. At each iteration a pair of agents are selected at random according to this distribution. For each such pair the inconsistency measure (definition \ref{inconsistency}) is evaluated and either both the valuations are replaced with the consensus valuation, or both are left unchanged, depending on the threshold $\gamma$ as in section \ref{sec:4}. The parameters for the simulation experiments are as follows:
	\begin{itemize}
		\item Population size: 100
		\item Language size: 5
		\item Initial beliefs: Random Boolean.
		\item Inconsistency threshold: $\gamma \in [0,1]$
	\end{itemize}

\begin{table}
\centering
\begin{tabular}{|c||c|c|}  \hline
binary operator & $0$ & $1$ \\ \hline \hline
$0$ & $0$ & $0:\frac{1}{2}, 1:\frac{1}{2}$ \\ \hline
$1$ & $0:\frac{1}{2}, 1:\frac{1}{2}$ & $1$ \\ \hline
\end{tabular}
\caption{The truth table  for the stochastic Boolean consensus operator}
\label{Booleanop}
\end{table}

Notice that here we are initialising the beliefs as random Boolean valuations (see section \ref{sec:4})\footnote{As a result of this Boolean initialisation, a language size of $5$ now produces a total of $2^5$ ($32$) possible valuations, as opposed to $3^5$ ($243$) possible valuations.}. This allows us to make a direct comparison between the performance of the three valued combination operator and a similar two valued operator. For the latter we assume that only binary truth values are available to represent an agent's beliefs. In this context, in order for two agents with conflicting truth values for $p_i$ (i.e. one $0$ and the other $1$) to reach consensus, we propose that they simply agree to pick one of the truth values at random e.g. by tossing a fair coin. Table \ref{Booleanop} gives the truth table for the operator in which directly conflicting truth values leads to a stochastic outcome.

The focus on simulations with $5$ propositional variables is intended to increase the number of opinions relative to the size of the population, in order to achieve a good distribution of valuations. For example, a language size of $5$ allows for $32$ possible Boolean valuations. With a population of $100$ agents, it is therefore very likely that each opinion will occur at least once. In comparison, a language size of $10$ produces $1,024$ possible Boolean valuations which severely decreases the probability of an opinion being present in a population of the same size.

\begin{figure*}[t]
	\centering
	\includegraphics[width=0.8\textwidth]{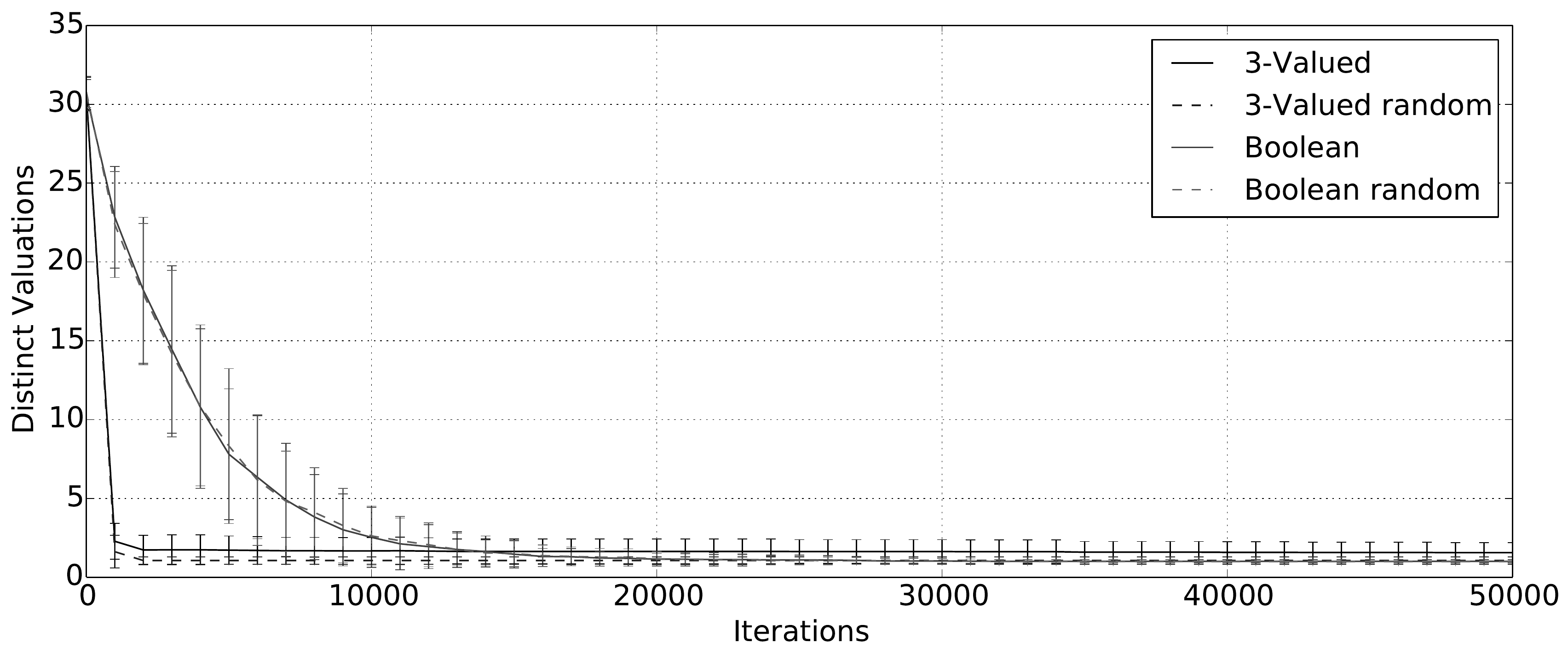}
	\caption{Trajectory showing the number of distinct valuations plotted against iterations for $\gamma=0.7$.}
	\label{fig:payoff_trajectory}
\end{figure*}

Figures \ref{fig:average_payoff}, \ref{fig:payoff_dkvs} and \ref{fig:payoff_trajectory} show the results for simulation experiments with agent selection based on payoff. The results shown are mean values with error bars taken over $100$ independent runs of the simulation. Figure \ref{fig:average_payoff} shows the average population payoff after $50,000$ iterations given as a percentage of the maximal possible payoff value i.e. the payoff for the valuation $(P,N)$ where $P=\{p_i:f(p_i)>0 \}$ and $N=\{p_i: f(p_i)<0\}$. For both the binary and the three valued operators we show results for simulations in which agents are selected according to payoff (three-valued, Boolean) and at random as in section \ref{sec:4} (three-valued random, Boolean random). We see that for all values of $\gamma$, the three valued operator with payoff based selection outperforms all of the other approaches. For the former we can also see that average payoff increases with $\gamma$. In contrast, for the other approaches, including the payoff operator with payoff based selection, the mean of the average population payoff remains close to $0$ after $50,000$ iterations. Figure  \ref{fig:payoff_dkvs} shows the mean number of distinct valuations across the population of agents after $50,000$ iterations. All four versions of the operators converge on a small set of shared beliefs for sufficiently large $\gamma$. For $\gamma\geq 0.4$ the mean number of distinct valuations is less than $5$ while for $\gamma \geq 0.8$ it is $1$. Figure  \ref{fig:payoff_trajectory} shows a trajectory of how the number of distinct valuations varies with each iteration when $\gamma=0.7$. We can see that both the three-valued models converge quickly (in just over $2000$) iterations while the Boolean models require considerably longer to converge (over $20,000$ iterations).

\section{Conclusions}
\label{sec:6}

In this paper we have explored the use of Kleene's three valued logic as a framework in which to model multi-agent consensus formation. We have proposed a three valued combination operator, the intuition behind which is that conflicting binary truth values are replaced with a borderline (middle) truth value.
A number of simulation experiments have been presented employing this operator.
These can be divided into two main categories. For the first type of experiments, agents are selected at random from the population and form a consensus valuation providing that the level of inconsistency of their respective opinions is below a threshold parameter $\gamma$. Otherwise they do not form a consensus and instead retain their current opinions. For these experiments we found that there is convergence to a smaller subset of shared opinions across the population. For higher $\gamma$ values there is convergence on average to a single shared opinion and furthermore this opinion is crisp i.e. it admits no borderlines. For intermediate values of $\gamma$ the system convergences to a small set of opinions which to some extent remain vague.

In the second type of experiments a payoff function over beliefs is introduced, and agents are selected for possible combination with probability proportional to the payoff value of their current beliefs. Here we compare the three value operator with a similar stochastic Boolean operator. We find that the three valued operator with payoff based agent selection results in convergence to a smaller shared set of beliefs  with significantly higher average payoff than that of the initial population. The Boolean operator does not perform well in this context and does not result in a significant increase in average payoff, which instead remains close to $0$ after $50,000$ iterations.

The results of the payoff based experiments show how a three valued model for consensus provides a number of improvements over a traditional Boolean model.
Firstly, we have shown that the introduction of Kleene valuations to capture the inherent vagueness of propositions does not, in the long run, lead to the mass adoption of borderline truth values as a result of conflict occurring in the population.
Instead, we have seen how vagueness is reduced at lower $\gamma$ values, and at higher $\gamma$ values the population converges towards completely crisp opinions on average, admitting no borderline cases.
In addition to this, we can see that the introduction of a payoff based model drives consensus towards those valuations which result in higher payoff on average. By selecting pairs of agents based on their perceived success, we can achieve an increase to overall payoff in a small number of iterations, compared to no significant increase in payoff for the Boolean model.
Therefore, we have shown that the three valued approach incorporating a payoff model can drive convergence across the population towards more successful opinions.

We suggest that the experiments presented in this paper show the potential of using three valued logic in consensus modelling. There is also significant scope to extend the research presented in several new directions. For example, the above studies concern consensus defined at the level of propositional variables. However, in many cases agents will be most concerned to reach agreement about a relevant set of compound statements. For example, they may need to reach agreement about a particular set of conditional statements, or equivalences. Hence, an important question is that of how best to extend our proposed consensus model so as to be applicable to compound logical expressions. Another significant question concerns uncertainty. Suppose that in addition to vagueness agents also quantify their uncertainty about beliefs. \cite{lawrydubois} propose an extension of the three valued framework in which agents' beliefs are represented by a probability distribution over Kleene valuations. Ongoing research concerns the design of simulation studies in which to evaluate the convergence and payoff based performance of this extended model.
Finally, it would be interesting to consider extensions to the operator which allows for consensus between groups rather than just pairs of agents.

\section*{Acknowledgements}
\label{sec:7}
This research is partially funded by an EPSRC PhD studentship as part of a doctoral training partnership (grant number EP/L504919/1).

\hspace{1em}

\noindent All underlying data is included in full within this paper.


\bibliographystyle{abbrv}
\bibliography{ssmcs_2015_arxiv}


\end{document}